\documentclass{article}

\usepackage{arxiv}

\usepackage[utf8]{inputenc} 
\usepackage[T1]{fontenc}    
\usepackage{hyperref}       
\usepackage{url}            
\usepackage{booktabs}       
\usepackage{amsfonts}       
\usepackage{nicefrac}       
\usepackage{microtype}      
\usepackage{lipsum}		
\usepackage{graphicx}
\usepackage{natbib}
\usepackage{doi}
\usepackage{multirow}

\title{Automatic Labels are as Effective as Manual Labels in Biomedical Images Classification with Deep Learning}

\author{
Niccol\`{o} Marini$^{1}$*,
\and
Stefano Marchesin$^{2}$*,
\and
Lluis Borras Ferris$^{1}$,
\and
Simon P\"{u}ttmann$^{3}$,
\and
Marek Wodzinski$^{1,4}$,
\and
Riccardo Fratti$^{1}$,
\and
Damian Podareanu$^{5}$,
\and
Alessandro Caputo$^{6,7}$,
\and
Svetla Boytcheva$^{8,9}$,
\and
Simona Vatrano$^{7}$,
\and
Filippo Fraggetta$^{7}$,
\and
Iris Nagtegaal$^{10}$,
\and
Gianmaria Silvello$^{2}$,
\and
Manfredo Atzori$^{1,11}$
\and
Henning M\"{u}ller$^{1,12}$
}

\begin{document}

\maketitle

\noindent
$^{1}$Information Systems Institute, University of Applied Sciences Western Switzerland (HES-SO Valais), Sierre, Switzerland\\
$^{2}$Department of Information Engineering, University of Padua, Padua, Italy\\
$^{3}$University of Applied Sciences and Arts Dortmund, Dortmund, Germany\\
$^{4}$Department of Measurement and Electronics, AGH University of Kraków, Krakow, Poland\\
$^{5}$SURFsara, Amsterdam, The Netherlands\\
$^{6}$Department of Pathology, Ruggi University Hospital, Salerno, Italy\\
$^{7}$Pathology Unit, Gravina Hospital Caltagirone ASP, Catania, Italy\\
$^{8}$Ontotext, Sofia, Bulgaria\\
$^{9}$Institute of Information and Communication Technologies, Bulgarian Academy of Sciences, Sofia, Bulgaria\\
$^{10}$Department of Pathology, Radboud University Medical Center, Nijmegen, The Netherlands\\
$^{11}$Department of Neurosciences, University of Padua, Padua, Italy\\
$^{12}$Medical faculty, University of Geneva, 1211 Geneva, Switzerland\\
$^{*}$Both authors contributed equally to this work.\\

\renewcommand{\shorttitle}{Automatic Labels are as Effective as Manual Labels in Biomedical Images Classification}

\hypersetup{
pdftitle={Automatic Labels are as Effective as Manual Labels in Biomedical Images Classification with Deep Learning},
pdfsubject={q-bio.NC, q-bio.QM},
pdfauthor={Niccol\`{o} Marini},
pdfkeywords={Automatic Weak Labels, Deep Learning, Histopathology Image Classification, Noisy Labels},
}

\begin{abstract}
\textbf{Background}: 
The increasing availability of biomedical data is helping to design more robust deep learning (DL) algorithms to analyze biomedical samples.
Currently, one of the main limitations to train DL algorithms to perform a specific task is the need for medical experts to label data. 
Automatic methods to label data exist, however automatic labels can be noisy and it is not completely clear when automatic labels can be adopted to train DL models. 

\textbf{Method}: 
This paper aims to investigate under which circumstances automatic labels can be adopted to train a DL model on the classification of Whole Slide Images (WSI).
The analysis involves multiple architectures, such as Convolutional Neural Networks (CNN) and Vision Transformer (ViT), and over 10’000 WSIs, collected from three use cases: celiac disease, lung cancer and colon cancer, which one including respectively binary, multiclass and multilabel data.

\textbf{Results}: 
The results allow identifying 10\% as the percentage of noisy labels that lead to train competitive models for the classification of WSIs.
Therefore, an algorithm generating automatic labels needs to fit this criterion to be adopted.
The application of the Semantic Knowledge Extractor Tool (SKET) algorithm to generate automatic labels leads to performance comparable to the one obtained with manual labels, since it generates a percentage of noisy labels between 2-5\%. 

\textbf{Conclusions}: 
Automatic labels are as effective as manual ones, reaching solid performance comparable to the one obtained training models with manual labels.

\end{abstract}


\section{Introduction}

\label{sec:intro}

\subsection{Background}

The development of deep learning (DL) algorithms is fostering the design of new tools that can be trained on clinical data without the need for human intervention, especially in domains where the cost of annotations is high, such as histopathology.
Histopathology is the gold standard to diagnose cancer \citep{VLC2021, DAT2021}. 
The domain involves the analysis of small tissue slices, to identify microscopic findings related to dangerous diseases \citep{GBC2009}, such as cancer.
Tissue slices undergo microscopic examination by a medical expert named pathologist, who usually needs up to an hour per image to analyze a single sample \citep{KGW2013}.
Despite the increasing digitization of tissue samples, histopathological samples are still rarely analyzed exploiting digital aid in clinical practice \citep{FGZ2017, FLA2021}.
Digital pathology is a domain involving the management and digitization of tissue specimens, called Whole Slide Images (WSI).
WSIs are high-resolution images stored with a pyramidal format, to capture different magnification levels of details \citep{MeC2022}.
Usually, the highest resolution levels result in a spatial high-resolution of 0.25–0.5$\mu$m per pixel, which corresponds to an optical resolution of 20-40x.
WSIs are usually coupled with pathology reports. 
Pathology reports are semi-structured free-text documents containing information about the patient anamnesis, the tissue specimen type and the findings and observations identified by a pathologist during the tissue examination \citep{Hew2020, HRA2020}.
WSIs and reports are usually stored in the Laboratory Information System (LIS), easily enabling sample retrieval.
The increasing collection of biomedical samples is encouraging the design of automatic tools to analyze WSIs under the computational pathology domain \citep{VLC2021, MaL2016, LCV2022}.
Most of the computational domain algorithms are currently based on deep learning, such as CNNs (Convolutional Neural Networks) or ViT (Visual Transformers) \citep{XXK2023, CVF2023}.

Even if computational pathology algorithms show accurate and robust performance, in tasks such as WSI classification or segmentation, several challenges are still open, such as data labels \citep{MaL2016, CHG2019, VLC2021, APA2019, CCL2022}.
Data labels are required to train supervised learning algorithms.
However, the collection of labels is not trivial, considering both strong and weak annotations.
Even if strong labels (i.e. pixel-wise annotations) usually achieve the most accurate performance when used to train a deep learning model, they require a pathologist to analyze samples, which can be time-consuming, so often unfeasible \citep{KDW2020}.
Therefore, the research based on the analysis of WSIs is mostly based on the exploitation of weak (i.e. image-level) labels.
Weak labels are related to the global image, even if they originate from a region of the image including specific characteristics, such as cancer \citep{DZY2020}.
Weak labels are inherently more noisy than pixel-wise annotations, since the regions leading to a specific label may be a small percentage of the whole image (e.g. 1-2\%).
For this reason, algorithms based on weak labels require larger training datasets to reach accurate performance.
Currently, most of weakly-supervised algorithms in computational pathology are based on Multiple Instance Learning (MIL) framework \citep{CCG2018}, which models the whole image as a bag of instances, where only the annotations about the global image are available.
MIL framework includes several algorithms, which lately showed high performance when adopted on large-scale datasets \citep{CHG2019, ITM2018, WLM2019, LWC2021, HFK2020, CCL2022}.
For example, \cite{CHG2019} showed that it is possible to reach almost perfect predictions on binary classification (cancer vs. non-cancer) using around 10’000 weakly-annotated WSIs, on three use cases: skin, breast, and prostate images.
The production of weak labels is faster than the strong ones, since they can be extracted from reports.
For example, the analysis of a report may require approximately 30 seconds/1 minute, in comparison with the analysis of an image, which requires around an hour.
However, human intervention is usually still required to analyze reports, unless the Laboratory Information System (LIS), where the samples and corresponding reports are stored, has a specific structure to retrieve automatically data, according to the characteristics that can be used as labels.
Unfortunately, most LISs do not show this feature, since they are organized in heterogeneous ways.

Automatic methods for extracting concepts from reports and using them as weak labels already exist \citep{MMO2022}, but noisy characteristics of weak labels can make automatic labeling ineffective. 
This paper aims to investigate under which circumstances automatic labels (i.e. labels automatically generated by an algorithm) can be adopted to train deep learning models to alleviate the need for experts to annotate data.
In particular, the goal is to identify when the results achieved using this type of labels reach results comparable to the ones obtained using manual labels (i.e. labels produced by a medical expert), so that data included in LISs can be fully exploited to build more robust and accurate tools to diagnose diseases.
The characteristics investigated in the paper involve the percentage of wrongly automatic labels necessary to reach comparable performance obtained with manual labels, the nature of labels (i.e. binary, multiclass and multilabel) and the deep learning architecture (robust or less robust to noise).
Wrongly automatic labels are annotations that are automatically produced by an algorithm and do not match the ground truth (i.e. manually-made).

\subsection{Contribution}

The paper includes a comparison of deep learning architecture trained with automatic and manual labels on the classification of WSIs.
The comparison involves two sets of experiments: a controlled scenario and a real-case scenario.
In the controlled scenario, manual labels are randomly perturbed with different percentages of noise, simulating the output of an algorithm to generate automatic labels.
The random perturbation involves a modification on the labels: in the celiac disease use case (binary), labels are flipped; in the lung cancer use case (multiclass), a different class is assigned to a sample; in the colon cancer use case (multilabel), labels are modified so that one or more classes from the original label are flipped.
In the real-case scenario, the Semantic Knowledge Extractor Tool (SKET) \citep{MGM2022} is used to extract meaningful concepts from reports that are used as weak labels for the corresponding samples.

The analysis involves three tissue use cases, celiac disease, lung cancer and colon cancer, composing a training dataset with over 10’000 WSIs, used to train three deep learning architectures: CLAM \citep{LWC2021}, transMIL \citep{SBC2021} and Vision Transformer (ViT) \citep{CCL2022}.

\begin{figure}[!h]

\centering

\includegraphics[height=0.5\textheight]{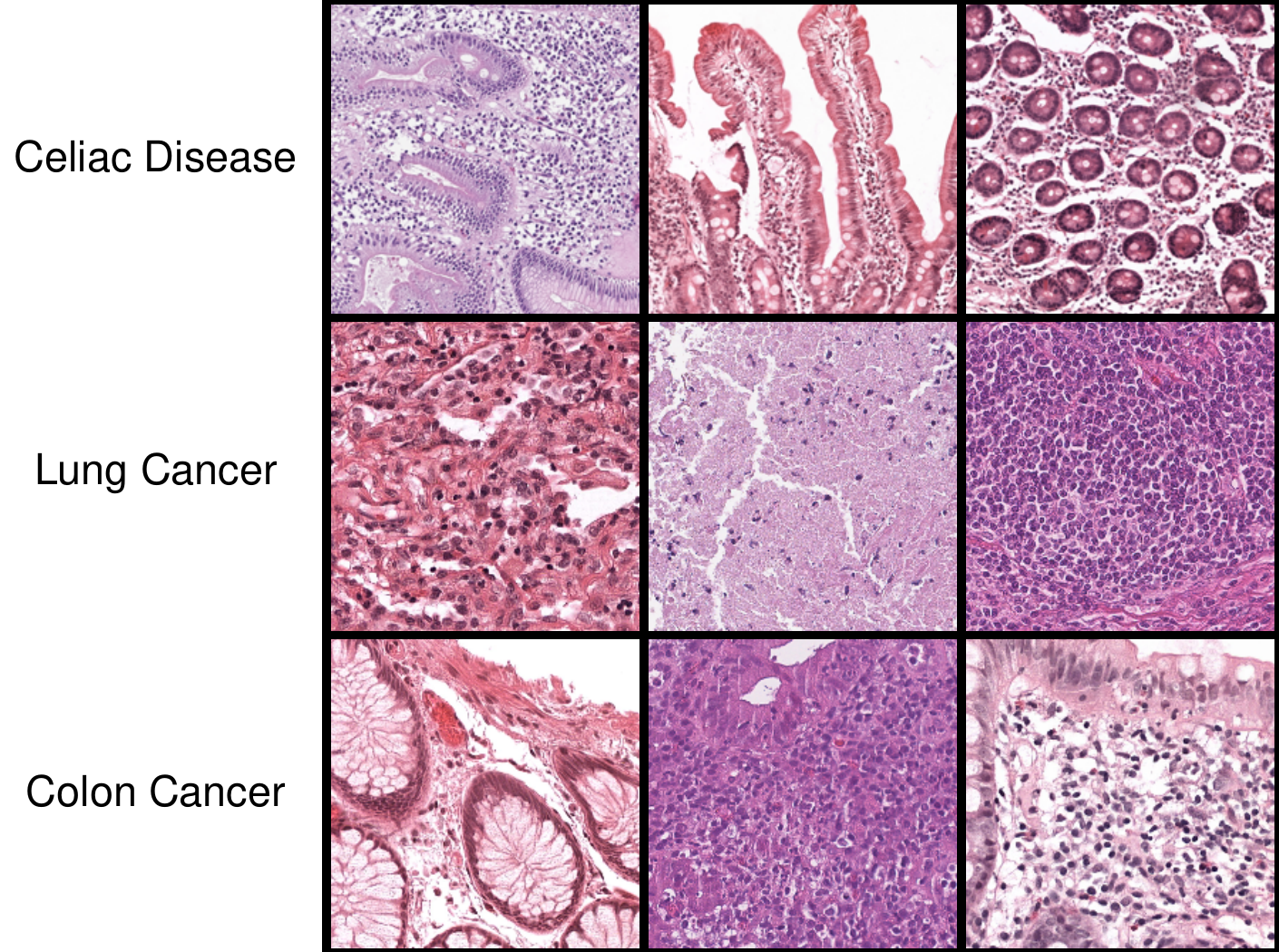}

\caption{Overview of the tissue use cases analyzed in the paper. The upper line includes examples of duodenal tissue samples, related to celiac disease. The central line includes examples of lung tissue samples. The bottom line includes examples of colon samples.} \label{fig:use_cases}

\end{figure}


Celiac disease (CD) is an autoimmune disorder leading to inflammations and damage in the small intestine, resulting in a range of gastrointestinal and systemic symptoms \citep{CVS2019}.
Globally, celiac disease affects about 1-2\% of the population \citep{LeB2021}, with variations across regions. 
Celiac disease diagnosis involves duodenal biopsies and serological tests (for specific antibodies).
In particular, the examination of biopsies aims to identify villous atrophy, crypt hyperplasia and increased intraepithelial lymphocytes.
In this paper, duodenal samples are labeled with celiac disease or normal tissue (binary labels).

Lung cancer is the leading cause of death related to cancer worldwide \citep{SoC2019, WHO2023}.
It is categorized into two main primary groups: Non-Small Cell Lung Cancer (NSCLC), which represents the large majority of cases (about 85\% of cases), and Small-Cell Lung Cancer (SCLC), which is less common, but more aggressive. 
Furthermore, NSCLC is further described with subtypes, such as LUng ADenocarcinoma (LUAD), LUng Squamous cell Carcinoma (LUSC).
Diagnosis of lung cancer through biopsies often involves the identification of irregular cell patterns, architectural distortion, and increased cellular density \citep{Tra2011}.
In this paper, lung samples are labeled with SCLC, LUAD, LUSC, Normal Tissue.

Colon cancer is the fourth most often diagnosed cancer worldwide \citep{BVA2018}. 
Colon cancer diagnosis involves the identification of multiple concepts, such as the presence of cancer and the evaluation of polyp shapes and possible abnormalities leading to dysplasia.
In this paper, colon samples are labeled with colon cancer, high-grade dysplasia (HGD), low-grade dysplasia (LGD), hyperplastic polyp and normal tissue (multilabel labels).
Figure \ref{fig:use_cases} shows some histopathological samples corresponding to the three tissues.

\section{Materials and Methods}

\subsection{Dataset composition}

The dataset used in this paper includes WSIs and reports (paired together) of celiac disease, lung cancer and colon cancer, collected from two hospitals: the Catania cohort and Radboudumc (RUMC).

WSIs are used to train and test different computer vision architectures on the image-level classification. 
WSIs are gigapixel tissue samples that can exhibit significant heterogeneity, for example in terms of staining \citep{MAO2021, MOW2023} and sample types. 
The image heterogeneity is a consequence of different acquisition procedures across laboratories, related to the chemical reagents applied to the specimen and to the whole slide scanners.
One of the main consequences of the heterogeneity is the stain variability, leading to different color variations, intensity, and uniformity of stains across different slides (as shown in Figure \ref{fig:use_cases}).
Also, the WSIs collected in this dataset show the same characteristic, aiming to replicate a common scenario in digital pathology.
WSIs collected from the Catania cohort were scanned with two 3DHistech scanners and two Aperio scanners and stored with a magnification of 20-40x; WSIs collected from RUMC were scanned using 3DHistech scanners, mainly stored at 40x magnification.

Reports are used to extract meaningful concepts, that are used as weak automatic labels to train the model to classify WSIs.
Reports include free-text descriptions summarizing the findings from tissue examination.
The findings are reported in a field named ‘Conclusion’, containing either macroscopic or microscopic observations.
Even if a report includes many fields, only the findings are relevant for the analysis proposed in the paper.
Therefore, additional patient information, such as family history or personal data, is discarded.
Textual reports show heterogeneity, mainly related to the source language and the textual content.
Reports are collected from an Italian and a Dutch hospital, therefore they have to be translated into English, to standardize the analysis.
The textual content slightly differs across sources, because the Catania cohort reports contain a field specifically for the findings identified in a single slide, while the RUMC reports include a specific field for the findings identified in a tissue block, which may encompass multiple slides.
Furthermore, samples are collected across years and are produced by many different pathologists, each one adopting its unique style of writing.

The dataset includes samples collected from three different use cases: celiac disease, lung cancer, colon cancer.
Data are randomly selected from LISs, to simulate a real-case scenario.
The goal is to show that the approach can generalize on different types of tissue (both in terms of images and reports).
Furthermore, different types of labels are used: celiac disease samples are annotated with binary labels, lung samples with multiclass labels, and colon samples with multilabel samples.

\begin{table}[!ht]
\centering
\footnotesize
\caption{Composition of the samples related to the celiac disease use case. Data are labeled with binary labels: celiac disease and normal tissue. The dataset is split into training and testing partitions. The model is trained and validated adopting a 10-fold cross-validation approach.}
\begin{tabular}{|cccc|}
\hline
\multicolumn{1}{|c|}{\textbf{Source}}     & \multicolumn{1}{c|}{\textbf{Celiac Disease}} & \multicolumn{1}{c|}{\textbf{Normal Tissue}} & \textbf{Total} \\ \hline
\multicolumn{4}{|c|}{\textbf{Training dataset: Automatic Labels}}                                                                            \\ \hline
\multicolumn{1}{|c|}{\textbf{Catania}}    & \multicolumn{1}{c|}{47}                      & \multicolumn{1}{c|}{711}                    & 758            \\ \hline
\multicolumn{1}{|c|}{\textbf{RUMC}} & \multicolumn{1}{c|}{217}                     & \multicolumn{1}{c|}{524}                    & 741            \\ \hline
\multicolumn{1}{|c|}{\textbf{Total}}      & \multicolumn{1}{c|}{264}                     & \multicolumn{1}{c|}{1235}                   & 1499           \\ \hline
\multicolumn{4}{|c|}{\textbf{Training dataset: Manual Labels}}                                                                                 \\ \hline
\multicolumn{1}{|c|}{\textbf{Catania}}    & \multicolumn{1}{c|}{61}                      & \multicolumn{1}{c|}{697}                    & 758            \\ \hline
\multicolumn{1}{|c|}{\textbf{RUMC}} & \multicolumn{1}{c|}{223}                     & \multicolumn{1}{c|}{518}                    & 741            \\ \hline
\multicolumn{1}{|c|}{\textbf{Total}}      & \multicolumn{1}{c|}{284}                     & \multicolumn{1}{c|}{1235}                   & 1499           \\ \hline
\multicolumn{4}{|c|}{\textbf{Testing dataset}}                                                                                                          \\ \hline
\multicolumn{1}{|c|}{\textbf{Catania}}    & \multicolumn{1}{c|}{10}                      & \multicolumn{1}{c|}{83}                     & 93             \\ \hline
\multicolumn{1}{|c|}{\textbf{RUMC}} & \multicolumn{1}{c|}{37}                      & \multicolumn{1}{c|}{63}                     & 100            \\ \hline
\multicolumn{1}{|c|}{\textbf{Total}}      & \multicolumn{1}{c|}{47}                      & \multicolumn{1}{c|}{146}                    & 193            \\ \hline
\end{tabular}
\label{tbl:celiac_data}
\end{table}

Table \ref{tbl:celiac_data} includes a detailed composition of data related to celiac disease collected from pathology reports, split into training and testing partitions. 
Data are labeled with binary labels: celiac disease and normal tissue.

\begin{table}[!ht]
\centering
\footnotesize
\caption{Composition of the samples related to the lung cancer use case. Data are labeled with multiclass labels: Small-Cell Cancer, Non-Small Adenocarcinoma Cell Cancer, Non-Small Squamous Cell Cancer, Normal Tissue. The dataset is split into training and testing partitions. The model is trained and validated adopting a 10-fold cross-validation approach.}
\begin{tabular}{|cccccc|}
\hline
\multicolumn{1}{|c|}{\textbf{Source}}     & \multicolumn{1}{c|}{\textbf{SCLC}} & \multicolumn{1}{c|}{\textbf{LUAD}} & \multicolumn{1}{c|}{\textbf{LUSC}} & \multicolumn{1}{c|}{\textbf{Normal}} & \textbf{Total} \\ \hline
\multicolumn{6}{|c|}{\textbf{Training dataset: Automatic Labels}}                                                                                                                                                                                        \\ \hline
\multicolumn{1}{|c|}{\textbf{Catania}}    & \multicolumn{1}{c|}{49}                  & \multicolumn{1}{c|}{526}                     & \multicolumn{1}{c|}{250}                    & \multicolumn{1}{c|}{226}             & 1051                                     \\ \hline
\multicolumn{1}{|c|}{\textbf{RUMC}} & \multicolumn{1}{c|}{1}                   & \multicolumn{1}{c|}{262}                     & \multicolumn{1}{c|}{195}                    & \multicolumn{1}{c|}{1041}            & 1499                                     \\ \hline
\multicolumn{1}{|c|}{\textbf{Total}}      & \multicolumn{1}{c|}{50}                  & \multicolumn{1}{c|}{788}                     & \multicolumn{1}{c|}{445}                    & \multicolumn{1}{c|}{1267}            & 2550                                     \\ \hline
\multicolumn{6}{|c|}{\textbf{Training dataset: Manual Labels}}                                                                                                                                                                                             \\ \hline
\multicolumn{1}{|c|}{\textbf{Catania}}    & \multicolumn{1}{c|}{50}                  & \multicolumn{1}{c|}{519}                     & \multicolumn{1}{c|}{271}                    & \multicolumn{1}{c|}{211}             & 1051                                     \\ \hline
\multicolumn{1}{|c|}{\textbf{RUMC}} & \multicolumn{1}{c|}{1}                   & \multicolumn{1}{c|}{260}                     & \multicolumn{1}{c|}{173}                    & \multicolumn{1}{c|}{1065}            & 1499                                     \\ \hline
\multicolumn{1}{|c|}{\textbf{Total}}      & \multicolumn{1}{c|}{51}                  & \multicolumn{1}{c|}{779}                     & \multicolumn{1}{c|}{444}                    & \multicolumn{1}{c|}{1276}            & 2550                                     \\ \hline
\multicolumn{6}{|c|}{\textbf{Testing dataset}}                                                                                                                                                                                                                      \\ \hline
\multicolumn{1}{|c|}{\textbf{Catania}}    & \multicolumn{1}{c|}{12}                  & \multicolumn{1}{c|}{62}                      & \multicolumn{1}{c|}{67}                     & \multicolumn{1}{c|}{32}              & 173                                      \\ \hline
\multicolumn{1}{|c|}{\textbf{RUMC}} & \multicolumn{1}{c|}{0}                   & \multicolumn{1}{c|}{55}                      & \multicolumn{1}{c|}{29}                     & \multicolumn{1}{c|}{110}             & 194                                      \\ \hline
\multicolumn{1}{|c|}{\textbf{Total}}      & \multicolumn{1}{c|}{12}                  & \multicolumn{1}{c|}{117}                     & \multicolumn{1}{c|}{96}                     & \multicolumn{1}{c|}{142}             & 367                                      \\ \hline
\end{tabular}
\label{tbl:lung_data}
\end{table}

Table \ref{tbl:lung_data} includes a detailed composition of data related to lung cancer collected from pathology reports, split in training and testing partitions. 
Data are labeled with multiclass labels: Small-Cell Cancer, Non-Small Adenocarcinoma Cell Cancer, Non-Small Squamous Cell Cancer, Normal Tissue.\\

\begin{table}[htb!]
\footnotesize
\centering
\caption{Composition of the samples related to the colon cancer use case. Data are labeled with multilabel annotations: Adenocarcinoma, High-Grade Dysplasia (HGD), Low-Grade Dysplasia (LGD), Hyperplastic Polyp, Normal Tissue. Due to the multilabel nature of labels, the total samples for each class may not correspond to the total number of samples. The dataset is split into training and testing partitions. The model is trained and validated adopting a 10-fold cross-validation approach.}

\begin{tabular}{|ccccccc|}
\hline
\multicolumn{1}{|c|}{\textbf{Source}}     & \multicolumn{1}{c|}{\textbf{Adenocarcinoma}} & \multicolumn{1}{c|}{\textbf{HGD}} & \multicolumn{1}{c|}{\textbf{LGD}} & \multicolumn{1}{c|}{\textbf{Hyperplastic}} & \multicolumn{1}{c|}{\textbf{Normal}} & \textbf{Total} \\ \hline
\multicolumn{7}{|c|}{\textbf{Training dataset: Automatic Labels}}                                                                                                                                                                                          \\ \hline
\multicolumn{1}{|c|}{\textbf{Catania}}    & \multicolumn{1}{c|}{776}                     & \multicolumn{1}{c|}{761}          & \multicolumn{1}{c|}{1288}         & \multicolumn{1}{c|}{511}                   & \multicolumn{1}{c|}{596}             & 3095           \\ \hline
\multicolumn{1}{|c|}{\textbf{RUMC}} & \multicolumn{1}{c|}{383}                     & \multicolumn{1}{c|}{377}          & \multicolumn{1}{c|}{853}          & \multicolumn{1}{c|}{943}                   & \multicolumn{1}{c|}{1341}            & 3460           \\ \hline
\multicolumn{1}{|c|}{\textbf{Total}}      & \multicolumn{1}{c|}{1159}                    & \multicolumn{1}{c|}{1138}         & \multicolumn{1}{c|}{2141}         & \multicolumn{1}{c|}{1454}                  & \multicolumn{1}{c|}{1937}            & 6555           \\ \hline
\multicolumn{7}{|c|}{\textbf{Training dataset: Manual Labels}}                                                                                                                                                                                               \\ \hline
\multicolumn{1}{|c|}{\textbf{Catania}}    & \multicolumn{1}{c|}{865}                     & \multicolumn{1}{c|}{774}          & \multicolumn{1}{c|}{1273}         & \multicolumn{1}{c|}{535}                   & \multicolumn{1}{c|}{570}             & 3095           \\ \hline
\multicolumn{1}{|c|}{\textbf{RUMC}} & \multicolumn{1}{c|}{394}                     & \multicolumn{1}{c|}{362}          & \multicolumn{1}{c|}{878}          & \multicolumn{1}{c|}{965}                   & \multicolumn{1}{c|}{1309}            & 3460           \\ \hline
\multicolumn{1}{|c|}{\textbf{Total}}      & \multicolumn{1}{c|}{1259}                    & \multicolumn{1}{c|}{1136}         & \multicolumn{1}{c|}{2151}         & \multicolumn{1}{c|}{1500}                  & \multicolumn{1}{c|}{1879}            & 6555           \\ \hline
\multicolumn{7}{|c|}{\textbf{Testing dataset}}                                                                                                                                                                                                                        \\ \hline
\multicolumn{1}{|c|}{\textbf{Catania}}    & \multicolumn{1}{c|}{111}                     & \multicolumn{1}{c|}{96}           & \multicolumn{1}{c|}{113}          & \multicolumn{1}{c|}{32}                    & \multicolumn{1}{c|}{98}              & 348            \\ \hline
\multicolumn{1}{|c|}{\textbf{RUMC}} & \multicolumn{1}{c|}{75}                      & \multicolumn{1}{c|}{65}           & \multicolumn{1}{c|}{146}          & \multicolumn{1}{c|}{119}                   & \multicolumn{1}{c|}{193}             & 520            \\ \hline
\multicolumn{1}{|c|}{\textbf{Total}}      & \multicolumn{1}{c|}{186}                     & \multicolumn{1}{c|}{161}          & \multicolumn{1}{c|}{259}          & \multicolumn{1}{c|}{151}                   & \multicolumn{1}{c|}{291}             & 868            \\ \hline
\end{tabular}
\label{tbl:colon_data}
\end{table}

Table \ref{tbl:colon_data} includes a detailed composition of data related to colon cancer collected from pathology reports, split into training and testing partitions. 
Data are labeled with multilabel labels: Adenocarcinoma, High-Grade Dysplasia (HGD), Low-Grade Dysplasia (LGD), Hyperplastic Polyp, Normal Tissue.

\subsection{Data analysis pipeline}

\begin{figure}[!h]
\centering
\includegraphics[width=1\linewidth]{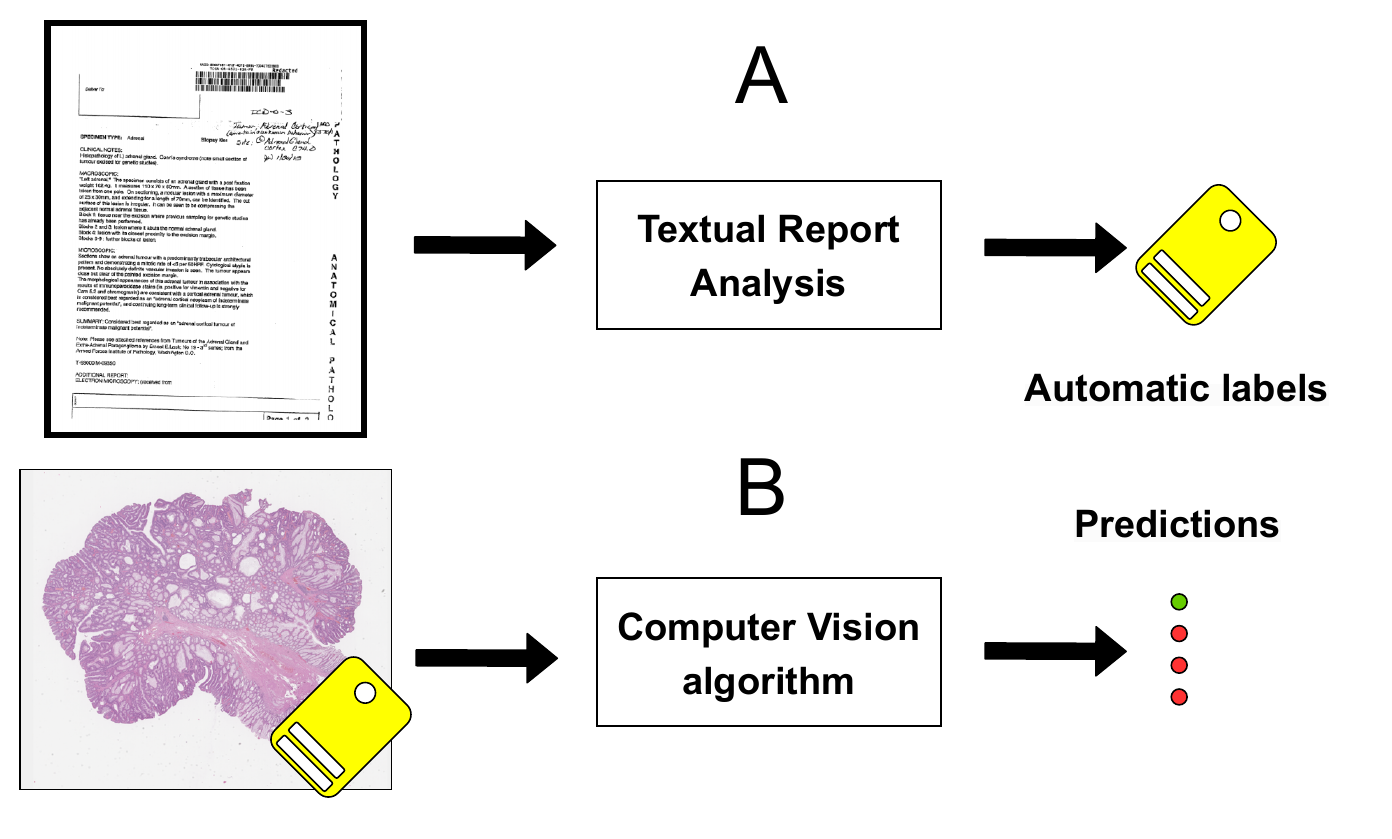}

\caption{Overview of the data analysis pipeline proposed in the paper. It includes two steps. The first step (A) involves the analysis of textual reports, to extract meaningful concepts that can be used as weak (automatic) labels for WSIs. The second step (B) involves image analysis through computer vision algorithms, that are transparent to the user and can be exchanged, to predict the content of the images. } \label{fig:pipeline}
\end{figure}

The training schema is based on computer vision algorithms to classify WSIs, comparing the performance of automatic and manual labels during the training.
Those algorithms are based on weak labels, since they are easier to be collected, even if they still require the intervention of medical experts.
In this paper, three different MIL backbones are adopted: two CNNs, CLAM and transMIL, and a ViT.
The architectures are trained to evaluate the effect that automatic labels may have on the training of models to classify WSIs.
Firstly, they are trained with noisy labels, randomly generated perturbating the manual labels, with a different percentage (1,2,5,10,20,50\%) of noise.
The goal of this experiment is to evaluate the effect that noisy labels have on the performance of a model.
However, this setup does not fit a real-case scenario where automatic labels are adopted.
Noisy labels may be considered as wrongly-labeled samples, but not all mistakes on labels have the same likelihood to happen.
Considering for example weak labels inferred by reports: some reports, due to their content, may be more easily mislabeled.
For this reason, a second setup is proposed, adopting a real tool to extract concepts from reports: the Semantic Knowledge Extractor Tool (SKET) \citep{MGM2022}.
This second experiment also helps to test the rules identified within the first experiment.

Figure \ref{fig:pipeline} shows an overview of the data analysis pipeline.

\subsection{Computer vision architectures}
Three computer vision algorithms to classify WSIs are compared in the paper as backbones, to evaluate the effect that noisy labels may have on different architectures, including two CNNs and a ViT.
The CNNs have a ResNet34 backbone, while the ViT has a backbone similar to the one shown in \cite{CCL2022}, considering a single magnification level.
In both cases, the backbones are designed to output an embedding of size 128 representing a single WSI, so that the same classifier can be adopted for all architectures, modifying the output classes based on the use case.

\paragraph{CLAM}
Clustering-constrained Attention Multiple Instance Learning (CLAM) \citep{LWC2021} is a MIL framework based on an attention-based network, whose goal is to highlight relevant regions inside the WSI, to improve the WSI-level prediction. 
CLAM exploits a mechanism on the single instances to aggregate them on clusters, according to the instance similarity, to enrich the WSI-representation and reach higher WSI-level predictions.
CLAM can have one or more attention branches, depending on the number of classes.
In this paper, a single attention branch (CLAM\_SB) is used when the model is used on celiac disease (binary labels), while instead a multiple attention branch (CLAM\_MB) is used on the other two use cases.

\paragraph{transMIL}
transMIL \citep{SBC2021} is a MIL framework developed to exploit the morphological and spatial characteristics of WSIs.
Even if morphological and spatial characteristics of images are important, the attention mechanism does not take them into account when evaluating input instances. 
transMIL exploits Transformer architectures \citep{VSP2017} to highlight relationships between single instances, modeling input instances as a sequence of tokens and evaluating the similarity among instances.

\paragraph{Vision Transformer}
Vision Transformer \citep{SNZ2021, HWC2020} is a deep learning architecture adopted to analyze images, adopting the self-attention mechanism to process input data instead of convolutional layers, showing more competitive performance in terms of accuracy and efficiency.
The architecture processes input data as a sequence of input tokens, that are small sub-regions of the input image (usually 16x16 pixels).
The architecture includes 12 encoder layers producing the embedding to feed the classifier.

\subsection{Semantic Knowledge Extractor Tool (SKET)}
SKET \citep{MGM2022} is an unsupervised algorithm combining a rule-based expert system with machine learning models, chosen to extract meaningful concepts from reports and use them as weak labels for WSIs  \citep{MGM2022, MSA2023}.
The algorithm includes three components: Named Entity Recognition, Entity Linking and Data Labeling.
Named Entity Recognition involves pre-trained models (ScispaCy models \citep{NKB2019}), developed to work on biomedical data, and large Word2Vec word vectors \citep{MSC2013} trained on the PubMed Central Open Access Subset \citep{MSC2013}.
Entity Linking involves a combination of similarity matching techniques to match ad-hoc concepts to a reference ontology.
Data Labeling involves the mapping of the concepts with a set of annotation classes.
SKET is an unsupervised model, therefore no training data are required to tune it.
This feature is relevant since it does not require data annotation for training, such as other Natural Language Processing (NLP) algorithms.

\section{Experimental Setup}

\subsection{Image pre-processing}
Image pre-processing includes the WSI splitting into patches.
WSIs usually do not fit modern GPU hardware memory because of their gigapixel characteristics, therefore they have to be split into patches.
In this paper, WSIs are split 224x224 pixel patches using the Multi\_Scale\_Tools library \citep{MOP2021}. 
The choice of the size is related to the characteristics of ResNet34 backbone, requiring fixed input size.
Patches are extracted from magnification 5x considering celiac samples, while lung and colon patches are sampled from magnification 10x. 
The magnifications are chosen considering that the magnification allows to identify peculiar morphological features, useful for the classification task: 5x for celiac disease cases, 10x for lung and colon ones.
The choice of the magnification to examine is driven by the characteristics of the problem to solve: celiac disease diagnosis requires to identify the villous shape and the crypts, therefore 5x magnification is chosen; on the other hand, lung and colon require a more refined level of magnification, because the shape of glands is as relevant as the cell infiltration, therefore 10x is chosen.
Not all sampled patches are selected: the ones from background regions are discarded, being not informative. 
The identification of background regions involves the application of HistoQC tool \citep{JZG2019}, which generates tissue masks.

\subsection{Report pre-processing}
The report pre-processing only involves their translation into English.
Original reports are stored in Italian and Dutch, depending on the workflow from which they are collected. 
The translation is necessary because state-of-the-art NLP algorithms are mostly developed to work with inputs in English. 
MarianMT neural machine translation models \citep{JGD2018} are used to translate the content of the reports to English. 

\subsection{Architecture pre-training}
The backbones of deep learning algorithms to analyze images are pre-trained using self-supervised algorithms: simCLR \citep{CKN2020} for the CNNs (CLAM and transMIL), DINO v2 \citep{ODM2023} for the ViT.

Both algorithms are adopted to learn meaningful features from unannotated input data, exploiting similarities and dissimilarities between input samples.
In this paper, the input data for the algorithms are the patches sampled from the training partition.
Since data are unannotated, no information is available regarding similarity among patches.
Therefore, data augmentation is adopted: samples are similar to their augmented versions and dissimilar from the other samples within a batch.
The algorithms differ on the data augmentation strategy.
simCLR is designed for CNNs and its augmentation pipeline includes several operations, applied with a probability of 0.5: random rotations (90/180/270 degrees), vertical/horizontal flipping, hue-saturation-contrast (HUE) color augmentation, RGB shift, color jitter, gaussian noise, elastic transformation, grid distortions. 
DINO is designed for ViT and involves a knowledge distillation mechanism: two networks are involved in the training, a teacher and a student.
The teacher is a larger model producing outputs that the student aims to mimic and replicate.
Both models are directly trained with two different augmented versions of input samples.
However, the student is also trained with a cropped version (96x96 pixels) of the teacher inputs.
DINO v2 augmentation pipeline includes two augmentation pipelines: the first one includes color jitter, horizontal/vertical flipping, gaussian blur; the second one includes color jitter, horizontal/vertical flipping, gaussian blur, solarization.

\subsection{Image data augmentation pipeline}
Albumentations library \citep{BIK2020} is adopted to apply data augmentation to input images.
The operations involved are random rotations (90/180/270 degrees), vertical/horizontal flipping and hue-saturation-contrast (HUE) color augmentation.  
The operations from the data augmentation pipeline are selected with a probability of 0.5 and applied at image-level, so that all the patches are augmented consistently

\subsection{Metric to evaluate the performance}
The performance of the models is evaluated in terms of WSI classification, using the weighted F1-score.
The classification problem can be defined as a binary problem (celiac disease), multiclass problem (lung cancer) or multilabel problem (colon cancer).
F1-score is a metric to measure the accuracy of a classifier, combining the recall and the precision. 
Precision evaluates how well a classifier is robust to avoid predicting negative samples as positive ones, while recall evaluates how well it correctly classifies all the positive samples.
In all the use cases, data may show unbalanced class distribution, since they are randomly selected from workflows, aiming to simulate a real-case scenario.
For this reason, weighted macro F1-score is adopted
Weighted F1-score tackles class imbalance, evaluating the F1-scores for the single classes and then averaging them according the class support (number of true samples for the class).
The weighted F1-score is reported in terms of average and standard deviation of the ten experiment repetitions, evaluated on the test partition.

\subsection{Statistical significance test}
The performance difference among different setups is evaluated through the Wilcoxon Rank-Sum test \citep{Woo2007}.
The test aims to establish if the results of two different experiments are statistically significantly different ($p$-value < 0.05).

\subsection{K-fold cross-validation}
All the setups presented in the paper are trained using $k$-fold cross-validation, in order to evaluate the robustness of the model on data used for training.
The training partition is divided into $k$ folders ($k$=10 in this paper).
During every training repetition, k-1 folders are used to train the model, while the other group is used to validate it.
Data are split in partitions considering the patients, so that WSIs collected from a patient cannot be in two different partitions

\subsection{Hardware and Software}
The experiments are developed exploiting Python libraries.
The deep learning algorithms are implemented and trained using PyTorch 2.2.0 and run on a Tesla V100 GPU. 
WSIs are accessed using openslide 3.4.1 \citep{GGH2013}. 
WSI pre-processing involves Multi\_Scale\_Tools library \citep{MOP2021} and data augmentation is applied using albumentations 1.3.1 \citep{BIK2020}. 
The performance of the model is quantitavely evaluated using the metrics implemented by scikit-learn 0.22. 

\subsection{Hyperparameters}
The optimal configuration setup of both CNN and ViT hyperparameters is identified using the grid search algorithm.
The optimal set is the one reaching the lowest loss function of the classification of WSIs, considering the validation partition.
The parameters tested with the grid search algorithm are: the batch size (4 selected; 1,2,4,8 tested); the CNN optimizer (Adam selected); the ViT optimizer (Adam selected; Adam, LARS and AdamW tested); the number of epochs when the CNN model is trained (15; over this number of epochs, the loss function evaluated on the validation partition no longer decreases); the number of epochs when the HIPT model is trained (15; over this number of epochs, the loss function evaluated on the validation partition no longer decreases); the learning rate ($10^{-4}$; $10^{-2}$, $10^{-3}$, $10^{-4}$, $10^{-5}$ were tested); the decay rate ($10^{-4}$; $10^{-2}$, $10^{-3}$, $10^{-4}$, $10^{-5}$ were tested); the number of nodes in the intermediate layer after the ResNet and the ViT backbone (128; 64, 128, 256, 512 were tested).

\section{Results}

\subsection{Automatic labels}

\begin{table}[h]
\centering
\caption{Overview of the performance reached by SKET on the extract of meaningful concepts from pathology reports, evaluated in terms of F1-score. The algorithm is evaluated considering the training partitions of three use cases (celiac disease, lung cancer, colon cancer), since SKET does not require any training. The performance is evaluated considering data from Catania, RUMC and their combination.}
\begin{tabular}{|c|c|c|c|}
\hline
\textbf{Use case}       & \textbf{Catania} & \textbf{RUMC} & \textbf{Cumulative} \\ \hline
\textbf{Celiac Disease} & 0.860            & 0.964               & 0.944               \\ \hline
\textbf{Lung Cancer}    & 0.969            & 0.975               & 0.976               \\ \hline
\textbf{Colon Cancer}   & 0.976            & 0.961               & 0.971               \\ \hline
\end{tabular}
\label{tbl:sket_results}
\end{table}

Meaningful concepts can be extracted from pathology reports without the need for human intervention and can be adopted as weak labels, dramatically reducing the time needed to collect labels.

The performance of SKET (tool to extract weak labels from reports) is evaluated on the training partition of the three use cases, since SKET does not require any training, being a ruled-based algorithm.
The extracted concepts are compared with the manual labels, provided by medical experts.
Table \ref{tbl:sket_results} summarizes the results.
On every use case, SKET reaches a weighted F1-score over 0.944, considering the cumulative testing partition. 
On the single pathology workflows, the lowest performance is reached considering the Catania testing partition of celiac disease data (0.860).
Otherwise, the algorithm reaches high-level performance, always over 0.960 in terms of F1-score.

Being effective, SKET can be adopted to mine unlabeled datasets and to annotate large amounts of data, that can be used to train deep learning models.
When tested on a Tesla V100 GPU, SKET requires among 0.006 and 0.03 seconds to extract concepts from a report, depending on its length.
Considering the worst case scenario, the algorithm is still a thousand times faster than a human expert who needs in the best case scenario 30 s per report.
Therefore, the application of SKET leads to save 99.99\% of time required in comparison with human experts.
For instance, the weak labeling of 10'000 WSIs would require 300'000 seconds (around 83 hours, without breaks) for human experts, in the best case scenario; on the other hand, it would require 300 seconds (five minutes), in the worst case scenario, to SKET.

\subsection{Celiac disease}

The classification performance of multiple computer vision architectures trained with binary automatically annotated data to classify celiac disease WSIs is as effective as the performance reached by models using manually annotated data.

Tables \ref{tbl:res_celiac_fake} and \ref{tbl:res_celiac_sket} summarize the results.
The highest performance using manual labels is reached using a ViT architecture (F1-score = 0.914 $\pm$ 0.014 on the test partition), even if on the Catania partition transMIL shows the highest performance.
The results are still similar for the three architectures.

Table \ref{tbl:res_celiac_fake} shows the classification performance obtained using binary manual labels and noisy labels.
This experiment aims to investigate general rules for the adoption of automatic labels on the binary classification of WSIs.
Considering all the architectures, the performance is similar to the one obtained using manual labels, especially until 10\% of training samples are wrongly-annotated, the difference in terms of performance is not statistically significant.
When the percentage of wrongly-annotated training is 20\% (or more) the performance degrades and the difference, compared with manual labels, is statistically significant, suggesting this percentage of wrongly-annotated labels can be considered as a threshold for the adopting of automatic weak labels in a binary classification scenario.

\begin{table}[htb!]
\footnotesize
\centering
\caption{Results on the classification of celiac disease, in terms of F1-score. The performance is evaluated considering three computer vision architectures: CLAM, transMIL, ViT. The architectures are trained with manual weak binary labels and with noisy weak labels, randomly perturbated according to different percentages of noise. The percentage of noisy labels is reported in the 'Noisy Labels' column, while the accuracy of the labels is reported in terms of F1-score, 'F1 labels' column. The goal is to evaluate the effect that noisy weak labels have on the binary classification of WSIs. For every setup, the F1-score average and standard deviation of the classification performance are reported, considering the models trained with the 10-fold cross-validation. The setups where the difference is statistically significant in terms of performance (compared with the models trained with manual labels) are marked with an asterisk (*).
}
\begin{tabular}{|c|c|c|c|c|c|}
\hline
\textbf{Noisy Labels}         & \textbf{F1 Labels} & \textbf{Model} & \textbf{Catania}    & \textbf{RUMC} & \textbf{Cumulative} \\ \hline
\multirow{3}{*}{Manual} & \multirow{3}{*}{-}     & CLAM\_SB       & 0.958 $\pm$ 0.009   & 0.846 $\pm$ 0.023   & 0.900 $\pm$ 0.012   \\ \cline{3-6} 
                        &                          & transMIL       & 0.968 $\pm$ 0.009   & 0.850 $\pm$ 0.019    & 0.906 $\pm$ 0.010   \\ \cline{3-6} 
                        &                          & ViT            & 0.953 $\pm$ 0.011   & 0.877 $\pm$ 0.021   & 0.914 $\pm$ 0.014   \\ \hline
\multirow{3}{*}{1\%}    & \multirow{3}{*}{0.977}   & CLAM\_SB       & 0.954 $\pm$ 0.016   & 0.849 $\pm$ 0.024   & 0.900 $\pm$ 0.018   \\ \cline{3-6} 
                        &                          & transMIL       & 0.968 $\pm$ 0.009   & 0.864 $\pm$ 0.010   & 0.914 $\pm$ 0.007   \\ \cline{3-6} 
                        &                          & ViT            & 0.954 $\pm$ 0.014   & 0.896 $\pm$ 0.019   & 0.925 $\pm$ 0.010   \\ \hline
\multirow{3}{*}{2\%}    & \multirow{3}{*}{0.968}   & CLAM\_SB       & 0.951 $\pm$ 0.012   & 0.873 $\pm$ 0.021   & 0.911 $\pm$ 0.014   \\ \cline{3-6} 
                        &                          & transMIL       & 0.965 $\pm$ 0.011   & 0.853 $\pm$ 0.021   & 0.907 $\pm$ 0.010   \\ \cline{3-6} 
                        &                          & ViT            & 0.944 $\pm$ 0.017   & 0.877 $\pm$ 0.021   & 0.910 $\pm$ 0.013   \\ \hline
\multirow{3}{*}{5\%}    & \multirow{3}{*}{0.933}   & CLAM\_SB       & 0.951 $\pm$ 0.019   & 0.862 $\pm$ 0.019   & 0.905 $\pm$ 0.017   \\ \cline{3-6} 
                        &                          & transMIL       & 0.958 $\pm$ 0.012*  & 0.857 $\pm$ 0.018   & 0.905 $\pm$ 0.011   \\ \cline{3-6} 
                        &                          & ViT            & 0.938 $\pm$ 0.026   & 0.880 $\pm$ 0.026   & 0.910 $\pm$ 0.020   \\ \hline
\multirow{3}{*}{10\%}   & \multirow{3}{*}{0.909}   & CLAM\_SB       & 0.952 $\pm$ 0.013   & 0.862 $\pm$ 0.023   & 0.905 $\pm$ 0.017   \\ \cline{3-6} 
                        &                          & transMIL       & 0.953 $\pm$ 0.026*  & 0.838 $\pm$ 0.033   & 0.893 $\pm$ 0.027   \\ \cline{3-6} 
                        &                          & ViT            & 0.957 $\pm$ 0.014   & 0.860 $\pm$ 0.023   & 0.906 $\pm$ 0.014   \\ \hline
\multirow{3}{*}{20\%}   & \multirow{3}{*}{0.804}   & CLAM\_SB       & 0.922 $\pm$ 0.026*  & 0.819 $\pm$ 0.029   & 0.869 $\pm$ 0.023*  \\ \cline{3-6} 
                        &                          & transMIL       & 0.933 $\pm$ 0.024*  & 0.822 $\pm$ 0.013*  & 0.875 $\pm$ 0.016*  \\ \cline{3-6} 
                        &                          & ViT            & 0.925 $\pm$ 0.017*  & 0.834 $\pm$ 0.025*  & 0.879 $\pm$ 0.017*  \\ \hline
\multirow{3}{*}{50\%}   & \multirow{3}{*}{0.566}   & CLAM\_SB       & 0.537 $\pm$ 0.228*  & 0.450 $\pm$ 0.081*  & 0.490 $\pm$ 0.145*  \\ \cline{3-6} 
                        &                          & transMIL       & 0.765* $\pm$ 0.097* & 0.502* $\pm$ 0.02*  & 0.633 $\pm$ 0.041*  \\ \cline{3-6} 
                        &                          & ViT            & 0.440 $\pm$ 0.302*   & 0.459 $\pm$ 0.029*  & 0.480 $\pm$ 0.141*  \\ \hline
\end{tabular}
\label{tbl:res_celiac_fake}
\end{table}

Table \ref{tbl:res_celiac_sket} shows the comparison of automatic labels, generated with SKET, and manual labels.
The comparison among automatic and manual labels shows a F1-score equal to 0.944, suggesting that the algorithm should lead to performance similar to the one obtained with noisy labels when the percentage of mislabeled data is between 2\% and 5\%.
The results confirm the hypothesis, since the performance is slightly worse than the one obtained using manual labels, but the gap is not statistically significant (according to the Wilcoxon Rank-Sum test, comparing every setup to the one where manual labels are used), showing the effectiveness of automatic labels in a binary classification scenario.

\begin{table}[htb!]
\footnotesize
\centering
\caption{Results on the classification of celiac disease, in terms of F1-score. The performance is evaluated considering three computer vision architectures: CLAM, transMIL, ViT. The architectures are trained with automatic and manual weak binary labels, generated extracting meaningful concepts from the corresponding pathology report, using SKET algorithm. The performance of SKET is reported in the 'Train label' column. The goal is to evaluate the effectiveness of automatic labels on the binary classification of WSIs. For every setup, the F1-score average and standard deviation of the classification performance are reported, considering the models trained with the 10-fold cross-validation. The setups where the difference is statistically significant in terms of performance (compared with the models trained with manual labels) are marked with an asterisk (*).}
\begin{tabular}{|c|c|c|c|c|c|}
\hline
\textbf{Noisy Labels}                     & \textbf{F1 Labels}  & \textbf{Model} & \textbf{Catania}  & \textbf{RUMC} & \textbf{Cumulative} \\ \hline
\multirow{3}{*}{Automatic} & \multirow{3}{*}{0.944} & CLAM\_SB       & 0.948 $\pm$ 0.015 & 0.857 $\pm$ 0.017   & 0.901 $\pm$ 0.013   \\ \cline{3-6} 
                                    &                        & transMIL       & 0.960 $\pm$ 0.012 & 0.845 $\pm$ 0.017   & 0.900 $\pm$ 0.014   \\ \cline{3-6} 
                                    &                        & ViT            & 0.938 $\pm$ 0.023 & 0.889 $\pm$ 0.024   & 0.915 $\pm$ 0.015   \\ \hline
\multirow{3}{*}{Manual}      & \multirow{3}{*}{-}     & CLAM\_SB       & 0.958 $\pm$ 0.009 & 0.846 $\pm$ 0.023   & 0.900 $\pm$ 0.012     \\ \cline{3-6} 
                                    &                        & transMIL       & 0.968 $\pm$ 0.009 & 0.85 $\pm$ 0.019    & 0.906 $\pm$ 0.010   \\ \cline{3-6} 
                                    &                        & ViT            & 0.953 $\pm$ 0.011 & 0.877 $\pm$ 0.021   & 0.914 $\pm$ 0.014   \\ \hline
\end{tabular}
\label{tbl:res_celiac_sket}
\end{table}

\subsection{Lung cancer}

The classification performance of multiple computer vision architectures trained with multiclass automatically annotated data to classify lung cancer WSIs is as effective as the performance reached by models using manually annotated data.

Tables \ref{tbl:res_lung_fake} and \ref{tbl:res_lung_sket} summarize the results.
The highest performance using manual labels is reached using a ViT architecture (F1-score = 0.763 $\pm$ 0.012) on both test partitions, dramatically outperforming the other two architectures (CLAM reaches 0.674 $\pm$ 0.016, while transMIL reaches 0.696 $\pm$ 0.016).

Table \ref{tbl:res_lung_fake} shows the classification performance obtained using multiclass manual labels and noisy labels.
This experiment aims to investigate general rules for the adoption of automatic labels on the multiclass classification of WSIs.
Considering all the architectures, the performance is similar to the one obtained using manual labels, especially until 20\% of training samples are wrongly-annotated, the difference in terms of performance is not statistically significant.
When the percentage of wrongly-annotated training is 50\% the performance degrades and the difference, compared with manual labels, is statistically significant, suggesting this percentage of wrongly annotated labels can be considered as a threshold for the adoption of automatic weak labels in a multiclass classification scenario.

\begin{table}[htb!]
\footnotesize
\centering
\caption{Results on the classification of lung cancer, in terms of F1-score. The performance is evaluated considering three computer vision architectures: CLAM, transMIL, ViT. The architectures are trained with manual weak multiclass labels and with noisy weak labels, randomly perturbated according to different percentages of noise. The percentage of noisy labels is reported in the 'Noisy Labels' column, while the accuracy of the labels is reported in terms of F1-score, 'F1 labels' column. The goal is to evaluate the effect that noisy weak labels have on the multiclass classification of WSIs. For every setup, the F1-score average and standard deviation of the classification performance are reported, considering the models trained with the 10-fold cross-validation. The setups where the difference is statistically significant in terms of performance (compared with the models trained with manual labels) are marked with an asterisk (*).}
\begin{tabular}{|c|c|c|c|c|c|}
\hline
\textbf{Noisy Labels}                     & \textbf{F1 Labels} & \textbf{Model} & \textbf{Catania}   & \textbf{RUMC} & \textbf{Cumulative} \\ \hline
\multirow{3}{*}{Manual} & \multirow{3}{*}{-}       & CLAM\_MB          & 0.617 $\pm$ 0.027  & 0.717 $\pm$ 0.023   & 0.674 $\pm$ 0.016   \\ \cline{3-6} 
                                    &                          & transMIL          & 0.635 $\pm$ 0.024  & 0.745 $\pm$ 0.024   & 0.696 $\pm$ 0.016   \\ \cline{3-6} 
                                    &                          & ViT               & 0.705 $\pm$ 0.033  & 0.812 $\pm$ 0.02    & 0.763 $\pm$ 0.012   \\ \hline
\multirow{3}{*}{1\%}                & \multirow{3}{*}{0.991}   & CLAM\_MB          & 0.624 $\pm$ 0.022  & 0.725 $\pm$ 0.021   & 0.681 $\pm$ 0.014   \\ \cline{3-6} 
                                    &                          & transMIL          & 0.634 $\pm$ 0.042  & 0.756 $\pm$ 0.012   & 0.700 $\pm$ 0.020   \\ \cline{3-6} 
                                    &                          & ViT               & 0.697 $\pm$ 0.035  & 0.817 $\pm$ 0.018   & 0.762 $\pm$ 0.021   \\ \hline
\multirow{3}{*}{2\%}                & \multirow{3}{*}{0.98}    & CLAM\_MB          & 0.621 $\pm$ 0.034  & 0.721 $\pm$ 0.016   & 0.677 $\pm$ 0.018   \\ \cline{3-6} 
                                    &                          & transMIL          & 0.642 $\pm$ 0.033  & 0.739 $\pm$ 0.011   & 0.695 $\pm$ 0.017   \\ \cline{3-6} 
                                    &                          & ViT               & 0.698 $\pm$ 0.032  & 0.807 $\pm$ 0.026   & 0.757 $\pm$ 0.026   \\ \hline
\multirow{3}{*}{5\%}                & \multirow{3}{*}{0.957}   & CLAM\_MB          & 0.609 $\pm$ 0.035  & 0.715 $\pm$ 0.022   & 0.670 $\pm$ 0.021    \\ \cline{3-6} 
                                    &                          & transMIL          & 0.622 $\pm$ 0.050  & 0.743 $\pm$ 0.015   & 0.687 $\pm$ 0.026   \\ \cline{3-6} 
                                    &                          & ViT               & 0.699 $\pm$ 0.027  & 0.809 $\pm$ 0.029   & 0.758 $\pm$ 0.020    \\ \hline
\multirow{3}{*}{10\%}               & \multirow{3}{*}{0.907}   & CLAM\_MB          & 0.601 $\pm$ 0.037  & 0.690 $\pm$ 0.034   & 0.653 $\pm$ 0.027   \\ \cline{3-6} 
                                    &                          & transMIL          & 0.615 $\pm$ 0.029  & 0.739 $\pm$ 0.025   & 0.683 $\pm$ 0.023   \\ \cline{3-6} 
                                    &                          & ViT               & 0.699 $\pm$ 0.026  & 0.808 $\pm$ 0.018   & 0.757 $\pm$ 0.015   \\ \hline
\multirow{3}{*}{20\%}               & \multirow{3}{*}{0.822}   & CLAM\_MB          & 0.579 $\pm$ 0.060   & 0.725 $\pm$ 0.038   & 0.658 $\pm$ 0.042   \\ \cline{3-6} 
                                    &                          & transMIL          & 0.614 $\pm$ 0.039  & 0.743 $\pm$ 0.017   & 0.684 $\pm$ 0.018   \\ \cline{3-6} 
                                    &                          & ViT               & 0.702 $\pm$ 0.018  & 0.808 $\pm$ 0.015   & 0.759 $\pm$ 0.012   \\ \hline
\multirow{3}{*}{50\%}               & \multirow{3}{*}{0.561}   & CLAM\_MB          & 0.409 $\pm$ 0.087* & 0.528 $\pm$ 0.069*  & 0.477 $\pm$ 0.065*  \\ \cline{3-6} 
                                    &                          & transMIL          & 0.483 $\pm$ 0.055* & 0.566 $\pm$ 0.027*  & 0.537 $\pm$ 0.031*  \\ \cline{3-6} 
                                    &                          & ViT               & 0.576 $\pm$ 0.049* & 0.701 $\pm$ 0.040*  & 0.643 $\pm$ 0.038*  \\ \hline
\end{tabular}
\label{tbl:res_lung_fake}
\end{table}

Table \ref{tbl:res_lung_sket} includes the comparison of automatic labels and manual labels.
This comparison represents a real-case scenario of automatic data labeling, where automatic labels are generated by extracting concepts from reports.
The comparison among labels shows a F1-score equal to 0.976, suggesting that the algorithm should lead to performance similar to the one obtained in the previous experiment using 2\% and 5\%.
The results confirm the hypothesis, since the performance is slightly worse than the one obtained using manual labels, but the gap is not statistically significant (according to the Wilcoxon Rank-Sum test, comparing every setup to the one where manual labels are used).

\begin{table}[htb!]
\footnotesize
\centering
\caption{Results on the classification of lung cancer, in terms of F1-score. The performance is evaluated considering three computer vision architectures: CLAM, transMIL, ViT. The architectures are trained with automatic and manual weak multiclass labels, generated extracting meaningful concepts from the corresponding pathology report, using SKET algorithm. The performance of SKET is reported in the 'Train label' column. The goal is to evaluate the effectiveness of automatic labels on the multiclass classification of WSIs. For every setup, the F1-score average and standard deviation of the classification performance are reported, considering the models trained with the 10-fold cross-validation. The setups where the difference is statistically significant in terms of performance (compared with the models trained with manual labels) are marked with an asterisk (*).}
\begin{tabular}{|c|c|c|c|c|c|}
\hline
\textbf{Noisy Labels}            & \textbf{F1 Labels}  & \textbf{Model} & \textbf{Catania}  & \textbf{RUMC} & \textbf{Cumulative} \\ \hline
\multirow{3}{*}{Automatic} & \multirow{3}{*}{0.976} & CLAM\_MB       & 0.623 $\pm$ 0.031 & 0.705 $\pm$ 0.028   & 0.67 $\pm$ 0.020    \\ \cline{3-6} 
                           &                        & transMIL       & 0.620 $\pm$ 0.027 & 0.740 $\pm$ 0.027   & 0.686 $\pm$ 0.018   \\ \cline{3-6} 
                           &                        & ViT            & 0.682 $\pm$ 0.041 & 0.820 $\pm$ 0.014   & 0.756 $\pm$ 0.022   \\ \hline
\multirow{3}{*}{Manual}    & \multirow{3}{*}{-}     & CLAM\_SB       & 0.617 $\pm$ 0.027 & 0.717 $\pm$ 0.023   & 0.674 $\pm$ 0.016   \\ \cline{3-6} 
                           &                        & transMIL       & 0.635 $\pm$ 0.024 & 0.745 $\pm$ 0.024   & 0.696 $\pm$ 0.016   \\ \cline{3-6} 
                           &                        & ViT            & 0.705 $\pm$ 0.033 & 0.812 $\pm$ 0.020    & 0.763 $\pm$ 0.012   \\ \hline
\end{tabular}
\label{tbl:res_lung_sket}
\end{table}

\subsection{Colon cancer}

The classification performance of multiple computer vision architectures trained with multilabel automatically annotated data to classify colon cancer WSIs is as effective as the performance reached by models using manually annotated data.

Tables \ref{tbl:res_colon_fake} and \ref{tbl:res_colon_sket} summarize the results.
The highest performance using manual labels is reached using a ViT architecture (F1-score = 0.831 $\pm$ 0.009) on both test partitions, dramatically outperforming the other two architectures (CLAM reaches 0.773 $\pm$ 0.015, while transMIL reaches 0.791 $\pm$ 0.008).

Table \ref{tbl:res_colon_fake} shows the classification performance obtained using multilabel manual labels and noisy labels.
This experiment aims to investigate general rules for the adoption of automatic labels on the multilabel classification of WSIs.
Considering all the architectures, the performance is similar to the one obtained using manual labels, especially until 20\% of training samples are wrongly-annotated, the difference in terms of performance is not statistically significant.
When the percentage of wrongly-annotated training is 50\% the performance degrades and the difference, compared with manual labels, is statistically significant, suggesting this percentage of wrongly-annotated labels can be considered as a threshold for the adoption of automatic weak labels in a multilabel classification scenario.

\begin{table}[htb!]
\footnotesize
\centering
\caption{Results on the classification of colon cancer, in terms of F1-score. The performance is evaluated considering three computer vision architectures: CLAM, transMIL, ViT. The architectures are trained with manual weak multilabel labels and with noisy weak labels, randomly perturbated according to different percentages of noise. The percentage of noisy labels is reported in the 'Noisy Labels' column, while the accuracy of the labels is reported in terms of F1-score, 'F1 labels' column. The goal is to evaluate the effect that noisy weak labels have on the multilabel classification of WSIs. For every setup, the F1-score average and standard deviation of the classification performance are reported, considering the models trained with the 10-fold cross-validation. The setups where the difference is statistically significant in terms of performance (compared with the models trained with manual labels) are marked with an asterisk (*).}
\begin{tabular}{|c|c|c|c|c|c|}
\hline
\textbf{Noisy Labels}         & \textbf{F1 Labels}  & \textbf{Model} & \textbf{Catania}   & \textbf{RUMC} & \textbf{Cumulative} \\ \hline
\multirow{3}{*}{Manual} & \multirow{3}{*}{-}     & CLAM\_MB       & 0.761 $\pm$ 0.015  & 0.780 $\pm$ 0.017   & 0.773 $\pm$ 0.015   \\ \cline{3-6} 
                        &                        & transMIL       & 0.771 $\pm$ 0.015  & 0.807 $\pm$ 0.007   & 0.791 $\pm$ 0.008   \\ \cline{3-6} 
                        &                        & ViT            & 0.824 $\pm$ 0.016  & 0.837 $\pm$ 0.007   & 0.831 $\pm$ 0.009   \\ \hline
\multirow{3}{*}{1\%}    & \multirow{3}{*}{0.988} & CLAM\_MB       & 0.761 $\pm$ 0.018  & 0.776 $\pm$ 0.016   & 0.771 $\pm$ 0.015   \\ \cline{3-6} 
                        &                        & transMIL       & 0.772 $\pm$ 0.014  & 0.810 $\pm$ 0.009    & 0.793 $\pm$ 0.010   \\ \cline{3-6} 
                        &                        & ViT            & 0.827 $\pm$ 0.018  & 0.835 $\pm$ 0.005   & 0.831 $\pm$ 0.009   \\ \hline
\multirow{3}{*}{2\%}    & \multirow{3}{*}{0.978} & CLAM\_MB       & 0.745 $\pm$ 0.018  & 0.764 $\pm$ 0.019   & 0.757 $\pm$ 0.017   \\ \cline{3-6} 
                        &                        & transMIL       & 0.777 $\pm$ 0.019  & 0.807 $\pm$ 0.010   & 0.793 $\pm$ 0.012   \\ \cline{3-6} 
                        &                        & ViT            & 0.821 $\pm$ 0.019  & 0.837 $\pm$ 0.005   & 0.831 $\pm$ 0.009   \\ \hline
\multirow{3}{*}{5\%}    & \multirow{3}{*}{0.943} & CLAM\_MB       & 0.765 $\pm$ 0.018  & 0.771 $\pm$ 0.021   & 0.769 $\pm$ 0.018   \\ \cline{3-6} 
                        &                        & transMIL       & 0.766 $\pm$ 0.013  & 0.808 $\pm$ 0.009   & 0.790 $\pm$ 0.008    \\ \cline{3-6} 
                        &                        & ViT            & 0.819 $\pm$ 0.015  & 0.835 $\pm$ 0.008   & 0.828 $\pm$ 0.009   \\ \hline
\multirow{3}{*}{10\%}   & \multirow{3}{*}{0.898} & CLAM\_MB       & 0.767 $\pm$ 0.023  & 0.777 $\pm$ 0.019   & 0.774 $\pm$ 0.018   \\ \cline{3-6} 
                        &                        & transMIL       & 0.768 $\pm$ 0.017  & 0.805 $\pm$ 0.009   & 0.789 $\pm$ 0.010   \\ \cline{3-6} 
                        &                        & ViT            & 0.827 $\pm$ 0.015  & 0.836 $\pm$ 0.005   & 0.833 $\pm$ 0.008   \\ \hline
\multirow{3}{*}{20\%}   & \multirow{3}{*}{0.814} & CLAM\_MB       & 0.748 $\pm$ 0.026  & 0.757 $\pm$ 0.020   & 0.754 $\pm$ 0.019   \\ \cline{3-6} 
                        &                        & transMIL       & 0.772 $\pm$ 0.012  & 0.809 $\pm$ 0.010   & 0.793 $\pm$ 0.008   \\ \cline{3-6} 
                        &                        & ViT            & 0.822 $\pm$ 0.020  & 0.833 $\pm$ 0.003   & 0.829 $\pm$ 0.009   \\ \hline
\multirow{3}{*}{50\%}   & \multirow{3}{*}{0.587} & CLAM\_MB       & 0.697 $\pm$ 0.042* & 0.646 $\pm$ 0.086*  & 0.670 $\pm$ 0.056*   \\ \cline{3-6} 
                        &                        & transMIL       & 0.723 $\pm$ 0.027* & 0.720 $\pm$ 0.024*  & 0.721 $\pm$ 0.015*  \\ \cline{3-6} 
                        &                        & ViT            & 0.811 $\pm$ 0.016* & 0.804 $\pm$ 0.021*  & 0.807 $\pm$ 0.016*  \\ \hline
\end{tabular}
\label{tbl:res_colon_fake}
\end{table}

Table \ref{tbl:res_colon_sket} includes the comparison of automatic labels and manual labels.
This comparison represents a real-case scenario of automatic data labeling, where automatic labels are generated extracting concepts from reports.
The comparison among labels shows a F1-score equal to 0.971, suggesting that the algorithm should lead to performance similar to the one obtained in the previous experiment using 2\% and 5\%.
The results confirm the hyphotesis, since the performance are slightly worse than the one obtained using manual labels, but the gap is not statistically significant (according to Wilcoxon Rank-Sum test, comparing every setup to the one where manual labels are used).

\begin{table}[htb!]
\footnotesize
\centering
\caption{Results on the classification of colon cancer, in terms of F1-score. The performance is evaluated considering three computer vision architectures: CLAM, transMIL, ViT. The architectures are trained with automatic and manual weak multilabel labels, generated extracting meaningful concepts from the corresponding pathology report, using SKET algorithm. The performance of SKET is reported in the 'Train label' column. The goal is to evaluate the effectiveness of automatic labels on the multilabel classification of WSIs. For every setup, the F1-score average and standard deviation of the classification performance are reported, considering the models trained with the 10-fold cross-validation. The setups where the difference is statistically significant in terms of performance (compared with the models trained with manual labels) are marked with an asterisk (*).}
\begin{tabular}{|c|c|c|c|c|c|}
\hline
\textbf{Noisy Labels}            & \textbf{F1 Labels}  & \textbf{Model} & \textbf{Catania}  & \textbf{RUMC} & \textbf{Cumulative} \\ \hline
\multirow{3}{*}{Automatic} & \multirow{3}{*}{0.971} & CLAM\_MB       & 0.761 $\pm$ 0.014 & 0.771 $\pm$ 0.019   & 0.767 $\pm$ 0.016   \\ \cline{3-6} 
                           &                        & transMIL       & 0.759 $\pm$ 0.013 & 0.801 $\pm$ 0.004   & 0.783 $\pm$ 0.005   \\ \cline{3-6} 
                           &                        & ViT            & 0.813 $\pm$ 0.014 & 0.836 $\pm$ 0.008   & 0.826 $\pm$ 0.008   \\ \hline
\multirow{3}{*}{Manual}    & \multirow{3}{*}{-}     & CLAM\_MB       & 0.761 $\pm$ 0.015 & 0.780 $\pm$ 0.017   & 0.773 $\pm$ 0.015   \\ \cline{3-6} 
                           &                        & transMIL       & 0.771 $\pm$ 0.015 & 0.807 $\pm$ 0.007   & 0.791 $\pm$ 0.008   \\ \cline{3-6} 
                           &                        & ViT            & 0.824 $\pm$ 0.016 & 0.837 $\pm$ 0.007   & 0.831 $\pm$ 0.009   \\ \hline
\end{tabular}
\label{tbl:res_colon_sket}
\end{table}

\section{Discussion}
This paper evaluates the application of weak automatic labels to train computer algorithms on classification.

The application of automatic weak labels would dramatically reduce the time needed to collect samples to train algorithms for the analysis of biomedical data.
However, it is not clear under which conditions automatic labels can be adopted to train algorithms.

The results achieved in the paper show that automatic labels are as effective as manual ones, for the classification of WSIs.
The first experiments (where manual labels are compared to different percentages of noisy labels) allow to identify some patterns in the algorithm performance.
The noise introduced by mislabeled samples (inherently present within automatic labels) impacts the performance of the networks, in terms of accuracy and robustness.
The performance achieved using small percentages of noisy labels is still comparable to the ones achieved using the manual labels, until a fixed percentage of mislabeled data: 10\% regarding  celiac disease (binary labels) and 20\% regarding lung and colon cancer (respectively multiclass and multilabel labels). 
This performance decrease can be explained considering the different natures of labels. 
Mislabeled samples have a high impact on binary classification, since the label flipping leads to opposite results.
Annotation errors are disruptive also in multiclass labels, even if in this case the effect can be smoothed if the errors involve similar classes (already prone to uncertainty).
Another explanation for this gap can be identified in the training dataset size. 
Another relevant parameter to consider when automatic labels are applied is the size of the training dataset, since the effect of mislabeled samples on the training may be compensated by the other samples.
In this paper, the celiac disease training dataset includes around 1'000 samples, while instead the lung cancer dataset includes around 2'500 samples and the colon cancer one includes around 6'500.
In the celiac disease use case, when the percentage of mislabeled samples is 20\% or more, the performance of the architectures is no longer comparable with the one reached using manual labels when the percentage of mislabeled samples is 20\%.
This result suggests that automatic labels can be adopted when the algorithm used to generate them is accurate.
The effect of noisy labels can be also identified on the performance standard deviation: the higher the percentage of noisy labels, the more the three architectures show less robustness.

The architectures trained using automatic labels reach performance comparable (i.e. the performance difference is not statistically significant) with the one reached using manual labels.
The results obtained using SKET to generate automatic weak labels show that automatic weak labels can be used to train different architectures on the classification of WSIs.
The conditions identified using randomly perturbated noisy data are also tested on a real case scenario, where the automatic labels are generated using SKET, an NLP algorithm to extract meaningful concepts from pathology reports.
This set of experiments is necessary to show the application of automatic labels in a real-case scenario, where the likelihood of mislabeling a sample varies. 
For example, if weak labels are automatically extracted from a report, depending on the report content a sample has more chances to be mislabeled.
This characteristic does not apply on the randomly perturbated noisy samples, where every sample can be randomly mislabeled.

The fact that automatic labels are as effective as manual labels opens many perspectives for the computational pathology domain and for the biomedical domain in general.
Automatic labels limit the need for medical experts to annotate data, which can save up to 99.99\% of time otherwise needed to analyze reports in order to infer labels.
Therefore, a dataset including around 10'000 can be weakly-annotated in around five minutes.
Considering the fact that every year a large amount of biomedical data is produced and only a small percentage is annotated, this would allow to exploit a vast amount of data, that can be used to build more accurate and robust models, helping medical experts to diagnose diseases more effectively.

\section{Conclusions}
The application of automatic labels may help to exploit vast amounts of unlabeled biomedical samples to train more robust models, reducing by 99.99\% the time needed to collect weakly-annotated samples.
However, is still not clear when this kind of labels is effective.
This paper evaluates the performance of different percentages of noisy labels (1,2,5,10,20,50\%) and compares the results with the performance obtained by the same architectures, but using manual weak labels, provided by medical experts.
After the identification of some rules (e.g. training datasets with 10\% of mislabeled samples lead to performance comparable to the one obtained using manual labels), SKET, an algorithm to extract meaningful concepts from reports, is used to generate automatic weak labels.
The performance reached by the models trained with SKET labels is comparable (not statistically significant difference) to the one obtained with manual labels, showing the effectiveness of automatic labels.
The result can allow to annotate samples contained in hospitals without the need of human efforts, paving the way to more and more accurate algorithms.
The code including the implement of the computer vision algorithms to classify WSIs is publicly available on Github (https://github.com/ilmaro8/wsi\_analysis). 

\section*{Acknowledgments}

This project has received funding from the European Union's Horizon 2020 research and innovation programme under grant agreement No. 825292 (ExaMode, \url{htttp://www.examode.eu/}). 
\bibliographystyle{unsrtnat}
\bibliography{references}  






\end{document}